\documentclass[aps,preprint,superscriptaddress,amssymb]{revtex4}
\usepackage{amsmath,amssymb,setspace,graphicx,subfigure,epsfig}

\usepackage{amsfonts}

\usepackage{latexsym}

\usepackage{epsfig}
\usepackage{graphicx}
\usepackage{epstopdf}
\usepackage{subfigure}

\usepackage{color}

\newcommand\lsim{\mathrel{\rlap{\lower4pt\hbox{\hskip1pt$\sim$}}
    \raise1pt\hbox{$<$}}}
\newcommand\gsim{\mathrel{\rlap{\lower4pt\hbox{\hskip1pt$\sim$}}
    \raise1pt\hbox{$>$}}}
\def\bea{\begin{eqnarray}}
\def\eea{\end{eqnarray}}
\def\ba{\begin{array}}
\def\ea{\end{array}}
\def\bc{\begin{center}}
\def\ec{\end{center}}
\def\nn{\nonumber}

\begin{document}

\title{Light Higgs Scenario in BMSSM and LEP Precision Data}

\author{Kyu Jung Bae}
\affiliation{FPRD and Department of Physics, Seoul National University, Seoul, 151-747, Korea}
\author{Radovan Dermisek}
\affiliation{Department of Physics, IndianaUniversity, Bloomington, IN47405, USA }
\author{Doyoun Kim}
\author{Hyung Do Kim}
\author{Ji-Hun Kim}
\affiliation{FPRD and Department of Physics, Seoul National University, Seoul, 151-747, Korea}

\date{\today}

\begin{abstract}

In this Letter we consider very light Higgs fields in BMSSM(Beyond MSSM).
The spectrum below TeV scale is the same as the MSSM but the Higgs potential
is modified and is well described in terms of effective dimension five and six operators.
A correction from the BMSSM operators allows us to consider new parameter space of 
Higgs sector which is not allowed in the MSSM. It can be regarded as a constrained version of general 2 Higgs doublet model (2HDM) as long as Higgs sector is concerned.
We focus on the possibility that CP odd Higgs (A) mass is about 7 or 8 GeV
and charged Higgs mass is comparable to W mass.
At the same time one of the CP even Higgs (h) is light enough such that
h and A production at the Z pole is kinematically allowed.
The tension between forward backward asymmetry of bottom quark $A^b_{\rm FB}$ measured at LEP
and the Standard Model prediction can be ameliorated if bottom quark pair produced from
light CP even Higgs is taken into account.

\end{abstract}

\maketitle

\section{Introduction}

The Standard Model (SM) has been almost completed by the discovery of top quark at the Tevatron
and the only missing ingredient is the Higgs.
LEP I/II experiments were done mainly to discover Higgs but without success up to Higgs mass 114 GeV \cite{Amsler:2008zzb}.
Though it rules out small part of the parameter space for the SM Higgs sector,
it rules out most of natural parameter space for the minimal supersymmetric standard model (MSSM).

In the MSSM, the quartic coupling of Higgs in the potential is determined from measured gauge couplings and the light CP even Higgs mass has an upper bound of about 120 GeV
(which can be 130 or 135 GeV if stop mixing is maximal) \cite{Djouadi:2005gi} \cite{Djouadi:2005gj}.
However, this upper bound is achieved only when the stop mass is as heavy as 1 TeV 
which makes it difficult to understand the weak scale out of it.
This `little hierarchy problem' in the MSSM has been considered seriously for recent several years
and many possible extensions of the MSSM have been proposed \cite{Giudice:2006sn}.
Even within the framework of MSSM, it was shown that the boundary condition at high energy which provides negative stop mass squared can reduce the fine tuning for the electroweak symmetry breaking
\cite{Dermisek:2006ey}
and explicit model has been proposed
\cite{Dermisek:2006qj} \cite{Kobayashi:2006fh}.

As an extension of the MSSM, NMSSM (next to MSSM) is one of the most popular scenarios \cite{Nilles:1982dy}.
Gauge sector extensions also have been proposed \cite{Batra:2003nj} \cite{Casas:2003jx}.
Recently BMSSM (beyond MSSM) has been proposed as a frame to study possible operators
which can affect the Higgs sector 
\cite{Dine:2007xi}.
There are extra fields above TeV scale but these new states can be integrated out below TeV such that we still keep the spectrum of the MSSM below TeV down to the weak scale. These new TeV particles modify the conventional Higgs potential and can increase the Higgs mass in this setup
\cite{Dine:2007xi}
and also Higgs mixing angle can be significantly changed such that Higgs phenomenology
can be quite different from standard one
\cite{Kim:2009sy}.
Electroweak baryogenesis with the light stop in BMSSM has also been studied \cite{Bernal:2009hd}.

The LEP bound is applied to the SM Higgs and in principle it can be weaker in the MSSM, NMSSM
or BMSSM if the production or decay is very different from the SM.
There had been extensive studies on nonstandard decay of Higgs
which can happen if there is an extra light particle (e.g., a singlet of NMSSM)
and the decay of Higgs is not just $b \bar{b}$
\cite{Dermisek:2005gg}.

In this Letter, we extend the work in the framework of BMSSM \cite{Kim:2009sy} which alters the Higgs phenomenology (both production and decay) significantly.
We assume that all new states other than the MSSM fields are heavier than TeV
such that they are captured only through the effective higher dimensional operators
after integrating out them.
Thus the spectrum is the same as the MSSM below TeV.
In this framework we propose a possibility of light Higgs which might have been produced at LEP I/II
\cite{Dermisek:2008id} \cite{Dermisek:2008dq} \cite{Park:2006gk}.
If CP odd Higgs (A) is lighter than Upsilon (bottomonium), it can alter Higgs decay significantly.
In addition if light CP even Higgs (h) is lighter than Z boson such that $Z \to h A$
is kinematically allowed, $b \bar{b}$ produced from $h$ decay can affect the electroweak precision data measured at LEP. In this case the suppression of $Z \to Zh$ is possible with the aid of BMSSM operators.

The contents of the Letter is following.
Firstly, various Higgs search bounds are briefly reviewed to convince that light Higgs scenario proposed here is compatible with all the existing bounds.
Secondly, we take the sample points of the BMSSM which have interesting features 
and can have interesting implications for the electroweak precision data of LEP.
Thirdly, we attempt to soften the discrepancy between the SM predictions and the LEP precision measurements within the scenario.
Finally, we summarize and conclude.

\section{LEP Higgs search bounds}

For the SM Higgs, LEP bound on the Higgs mass is 114 GeV at 95 percent confidence level (C.L.).
This equally applies to the Higgs field which has the same coupling to Z boson
and decays in the same way as the SM Higgs. Therefore, there are two ways to avoid the bound.
If the production is suppressed, the bound becomes weaker.
The modification of decay also makes the bound weaker depending on the channel.
For the production, Higgsstrahlung is suppressed for light CP even Higgs if $ZZh$ coupling is small.
For $g^2_{ZZh}=0.04 g^{2 \ \rm SM}_{ZZh}$, the Higgs can be as light as 70 to 75 GeV
from decay mode independent search \cite{Abbiendi:2002in}.

In this case the other CP even Higgs $H$ couples to Z boson
with almost the same strength as the SM Higgs
since there is a sum rule, $g^2_{ZZh}+g^2_{ZZH} = g^{2 \ \rm SM}_{ZZh}$.
For $H$, we can modify its decay if CP odd Higgs decays mostly to $AA$ rather than to $b\bar{b}$.
If $Br(H \to b\bar{b}) \le 0.2$ and $m_A < 10$ GeV (lighter than $2m_b$), 
$H$ can be as light as 100 GeV 
as $H \to AA \to 4 \tau$ does not give a strong constraint.

Once $m_A$ is very light, we also expect the charged Higgs mass to be close to W boson mass 
in MSSM-like theories as the tree level mass relation between charged Higgs, CP odd Higgs and W boson is following.
\bea
m_{H^{\pm}}^2 & = & m_W^2 + m_A^2.
\eea
This is violated by loop corrections in the MSSM but the violation is very tiny.
In BMSSM, the modification can be sizable depending on which operators are added,
but still the charged Higgs will be at around the weak scale.
Such a light Higgs might be dangerous. However, it was shown that it can be perfectly consistent
with the charged Higgs search from the top decay at the Tevatron since the charged Higgs decays 
not only to $\tau \nu$ (and $c s$) but also decays to $A W^{\pm *}$
\cite{Dermisek:2008dq}.

If $h$ and $A$ production is kinematically allowed at the Z pole,
it can provide a very interesting signature.
For small $ZZh$ coupling which is needed to keep $h$ lighter than Z boson,
$ZAh$ coupling is almost maximal and $hA$ associated production is possible.
The branching ratio of Z to Higgs is typically a few percent of Z to $b \bar{b}$.
There are two possibilities. If $h$ mainly decays to $AA$,
$Z \to h A \to 3A \to 6\tau$ puts a bound on $h$ mass.
If $h$ is heavier than $70$ GeV, the scenario is consistent with the current search bound.
More interesting possibility is the case when $h$ mainly decays to $b\bar{b}$.
Although it requires a fine tuning in the parameter choice since $h$ coupling to $A$
is of order one while it couples to $b$ with a bottom Yukawa coupling which is smaller than one
except at large $\tan \beta$ ($\tan \beta \sim 60$).
In this case, $Z \to h A \to b \bar{b} \tau^+ \tau^-$ would be the main decay channel.
For $m_h \ge 70$ GeV, $h$ and $A$ are produced with very small kinetic energy
and two taus decayed from $A$ would be very soft.
Two tau jets carry about 15 GeV of energy (7.5 GeV each)
and the measured energy of each tau jet will be typically smaller than 5 GeV
since it emits at least one tau neutrino and can emit more in leptonic decays.
If the energy of tau jet is less than 5 GeV, it is too soft to be identified as tau jet
and the whole event will be recorded as hadron events.

In this Letter, we focus on the scenario in which $Z \to hA$ is kinematically allowed at the Z pole
and $h$ mainly decays to $b \bar{b}$. This happens for $m_h \sim 70$ GeV
and $m_A \le 2m_b$. To take into account the recent BaBar result on Upsilon decay,
we take $m_A = 7$ or $8$ GeV as a representative value. The change of $m_A$ does not affect the result very much other than the light CP odd Higgs search bound.

\section{BMSSM and sample points}

The above mentioned scenario is hard to be realized in the MSSM.
The eigenvalue of the light CP even Higgs is too small (less than 50 or 60 GeV)
if CP odd Higgs mass is below 10 GeV in the MSSM.
Also the light CP even Higgs usually couples more strongly to Z boson than the heavy one.
Therefore, it is not possible to satisfy the direct search bound in the MSSM for such a light CP odd Higgs.

The scenario can be realized in the BMSSM
if we include BMSSM operators.
If there are new particles at around 1 TeV or higher
and if they couple to Higgs fields with order one coupling,
we can generate effective dimension five and dimension six operators 
which can give corrections comparable to the usual D term quartic couplings of Higgs fields.
The BMSSM just adds new operators such that they can alter the mass and the couplings
of the Higgs fields but does not introduce new light states into which Higgs can decay.
Therefore, Higgs decay is modified in the BMSSM only through the modification of Higgs couplings, e.g., $h \to AA$.

There are many operators with effective dimension five and six in the BMSSM.
In this Letter, we just consider two of them which might be relevant to the discussion.
By including other operators, the whole parameter space would be expanded.
Nevertheless, the main feature of the scenario would be the same.

The operators are
\bea
\delta V & = & 2\epsilon_1 H_u H_d (H_u^{\dagger} H_u + H_d^{\dagger} H_d) + h.c. \nn \\
&& + \epsilon_2 (H_u H_d)^2 + h.c. \nn \\
&& + \epsilon_3 (H_u^{\dagger} H_u)^2 + \epsilon_4 H_u H_d (H_u H_d)^{\dagger}. 
\eea
$\epsilon_3$ and $\epsilon_4$ are real and $\epsilon_1$ is assumed to be real to simplify the discussion.
CP even Higgs mass matrix ${\cal M}^2$ is given as follows.
{\small
\bea
\left( \begin{array}{cc}  (M_Z^2+4v^2 \epsilon_2) \cos^2 \beta +m_A^2 \sin^2 \beta + 4 v^2 \epsilon_1 \sin 2 \beta 
&(-M_Z^2-m_A^2+2\epsilon_4 v^2) \sin \beta \cos \beta + 4v^2 \epsilon_1\\
(-M_Z^2-m_A^2+2\epsilon_4 v^2) \sin \beta \cos \beta + 4v^2 \epsilon_1 
&\{M_Z^2 +2v^2(2 \epsilon_2 +\epsilon_3)\} \sin^2 \beta +m_A^2 \cos^2 \beta + 4v^2 \epsilon_1 \sin 2 \beta  \end{array} \right) . \nn
\eea}
$\epsilon_1, \epsilon_2$ and $\epsilon_4$ are the operators that does not exist in the MSSM
but can arise in the BMSSM after integrating out massive states at TeV and/or with supersymmetry breaking.
$\epsilon_3$ exists already in the MSSM from top-stop loop but here we consider more general $\epsilon_3$ which is not directly related to the stop mass. $\epsilon_3$ can arise in the BMSSM if there is an extra U(1)
and only $H_u$ is charged under the extra U(1).
We do not discuss the detailed BMSSM model beyond TeV.
Instead we will focus on the effective operators.
$\epsilon_1$ affects both the diagonal elements and the off-diagonal elements.
$\epsilon_2$ enters only in the diagonal entries.
$\epsilon_4$ affects only the off-diagonal elements. The dependence of CP even Higgs mass on BMSSM parameters can be read off from the two by two matrix.
The one in the diagonal elements can increase the eigenvalues if it is positive.
On the other hand the cancelation in the off-diagonal elements can reduce the level repulsion
such that the lightest eigenvalue can be larger than before.
Thus as long as the light CP even Higgs mass is concerned, the role of $\epsilon_2$ and $\epsilon_4$ are almost the same. 

Tree level mass of Charged Higgs is given by
\bea
m_{H^\pm}^2 = m_A^2 + m_W^2 + 2\epsilon_2 v^2 + \epsilon_4 v^2.
\eea
$\epsilon_1$ and $\epsilon_3$ does not affect the relation between $m_A$ and $m_{H^\pm}$.
Sizable corrections to the MSSM relation between $m_A$ and $m_{H^\pm}$ are expected
if $\epsilon_2$ and/or $\epsilon_4$ are sizable.
As charged Higgs mass bound is stiff at around $m_W$, i.e., it can not be much lighter than W boson mass, $\epsilon_2$ and $\epsilon_4$ are expected to be positive if sizable.
In the following discussion, we mainly focus on the effects of $\epsilon_1$ and $\epsilon_2$. The role of $\epsilon_4$ is very similar to $\epsilon_2$ as long as light CP even Higgs mass and charged Higgs mass are concerned. Only the impact on heavy CP even Higgs mass will be different and the phenomenological distinction is less clear. $\epsilon_3$ is taken to be -0.1 in the following discussion.
It can be achieved for very light stop or can be obtained with BMSSM operators. The origin of the operator is not the main concern of this paper.

\begin{figure}[thb]
\subfigure[]{
\includegraphics[width=3.in]{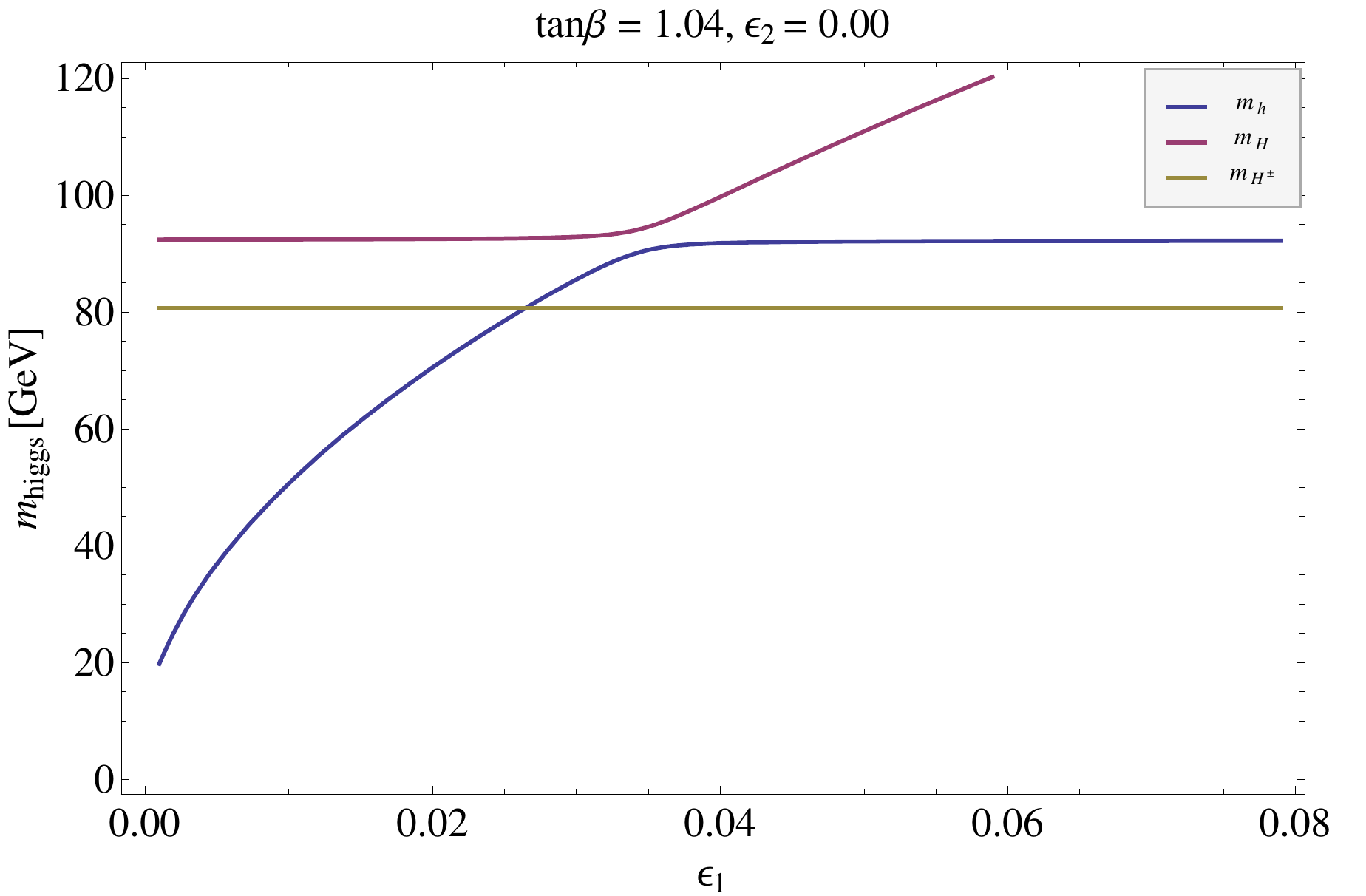}}
\subfigure[]{
\includegraphics[width=3.in]{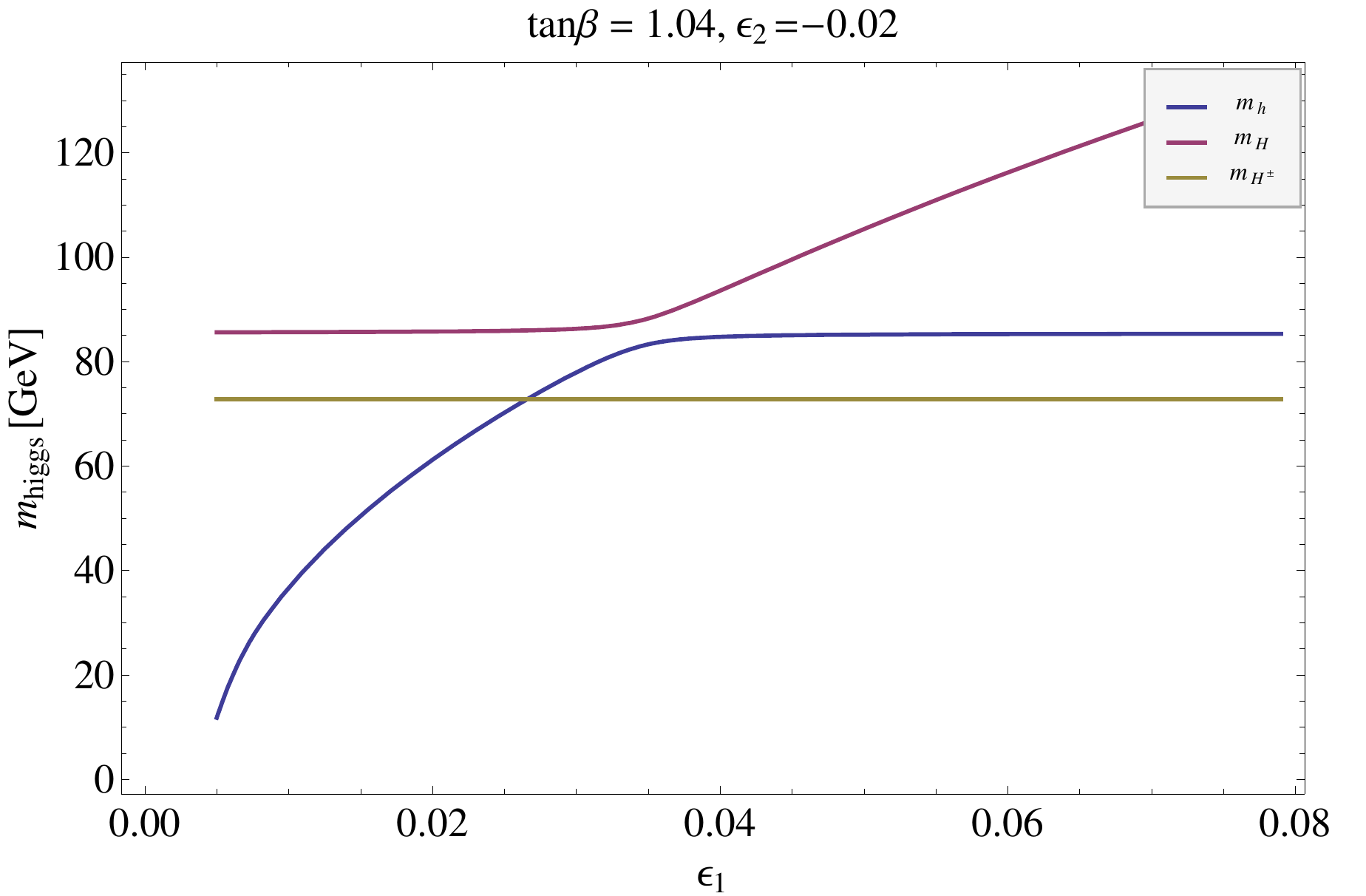}}
\subfigure[]{
\includegraphics[width=3.in]{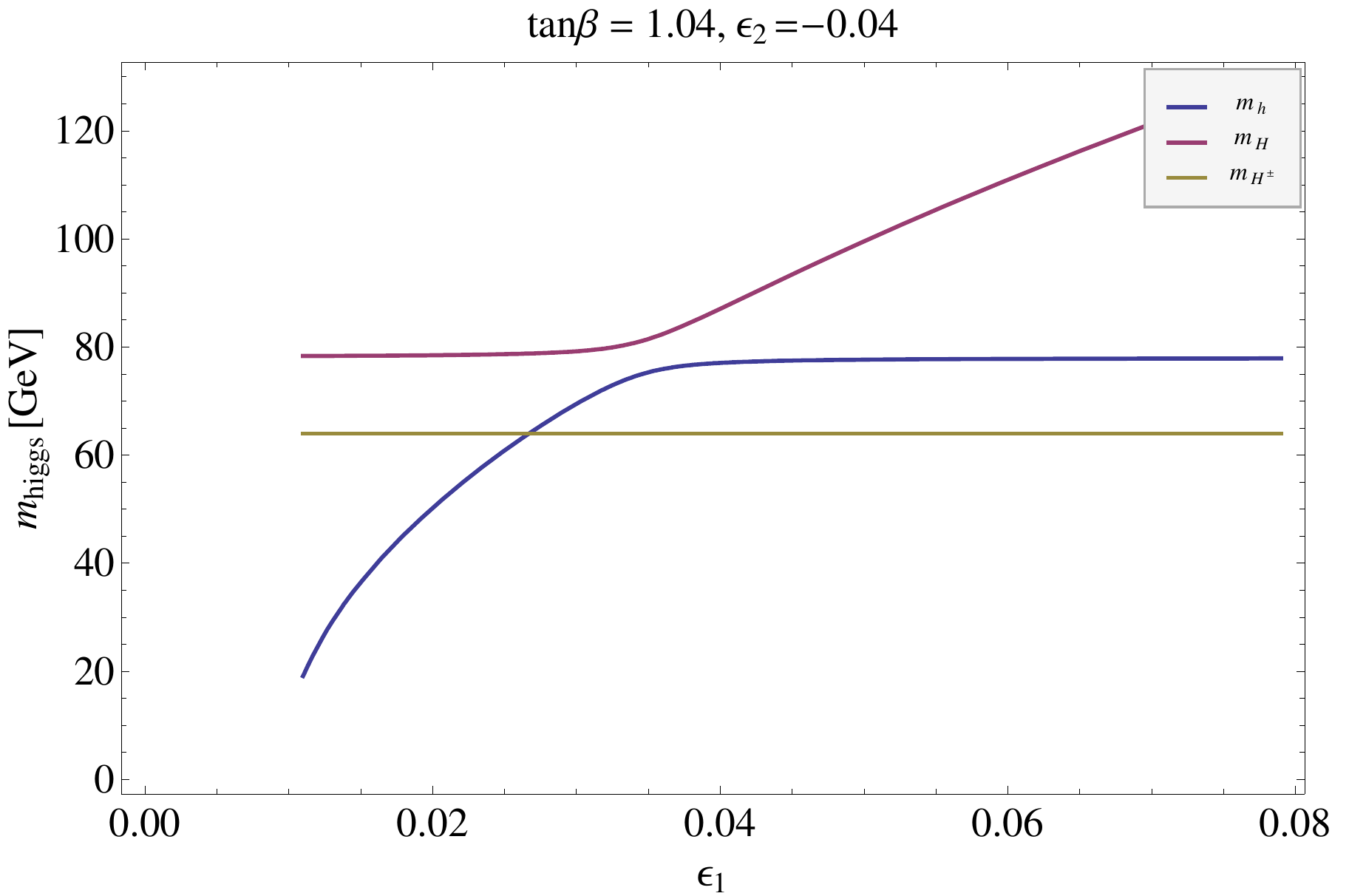}}
\subfigure[]{
\includegraphics[width=3.in]{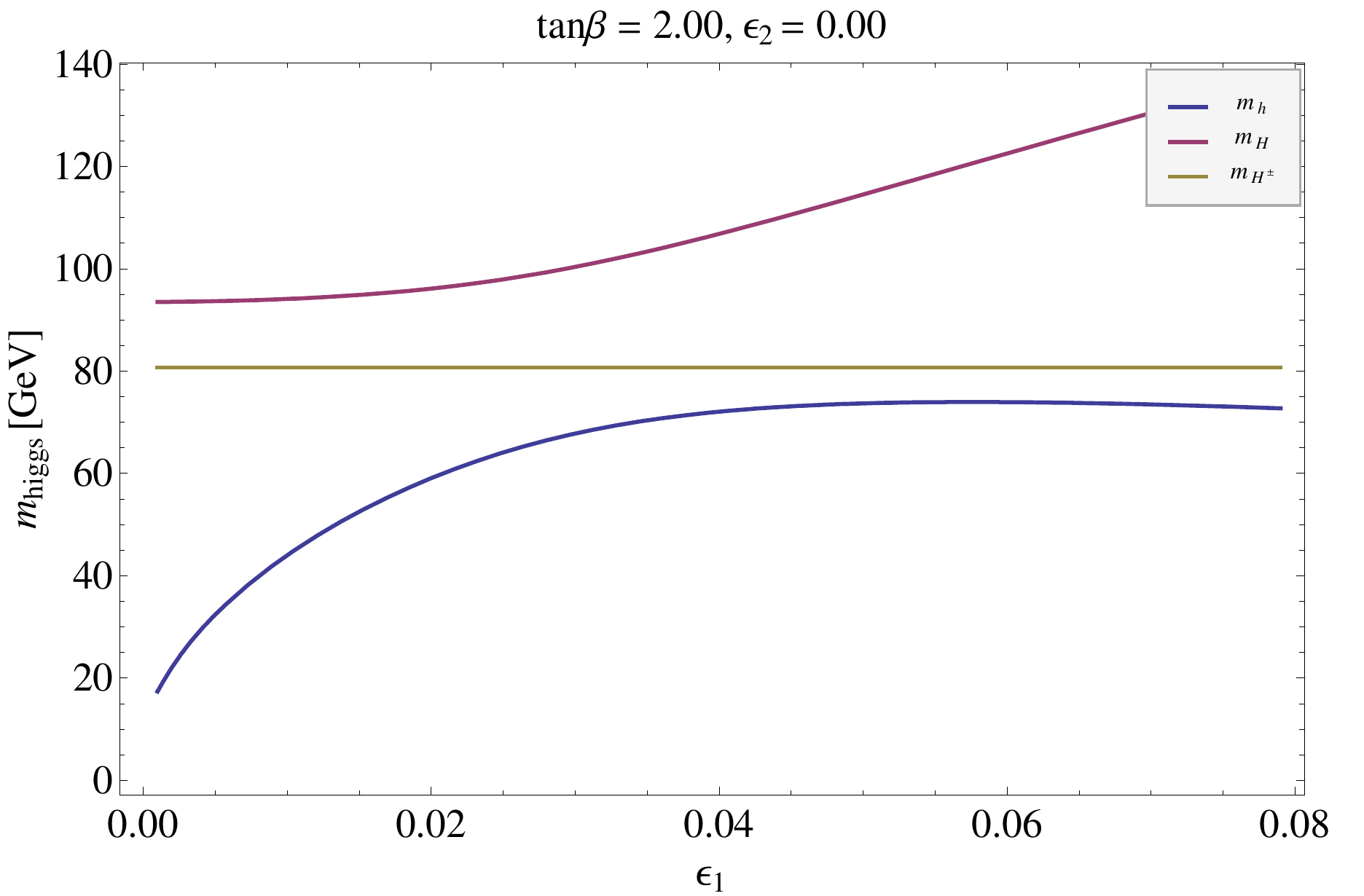}}
\subfigure[]{
\includegraphics[width=3.in]{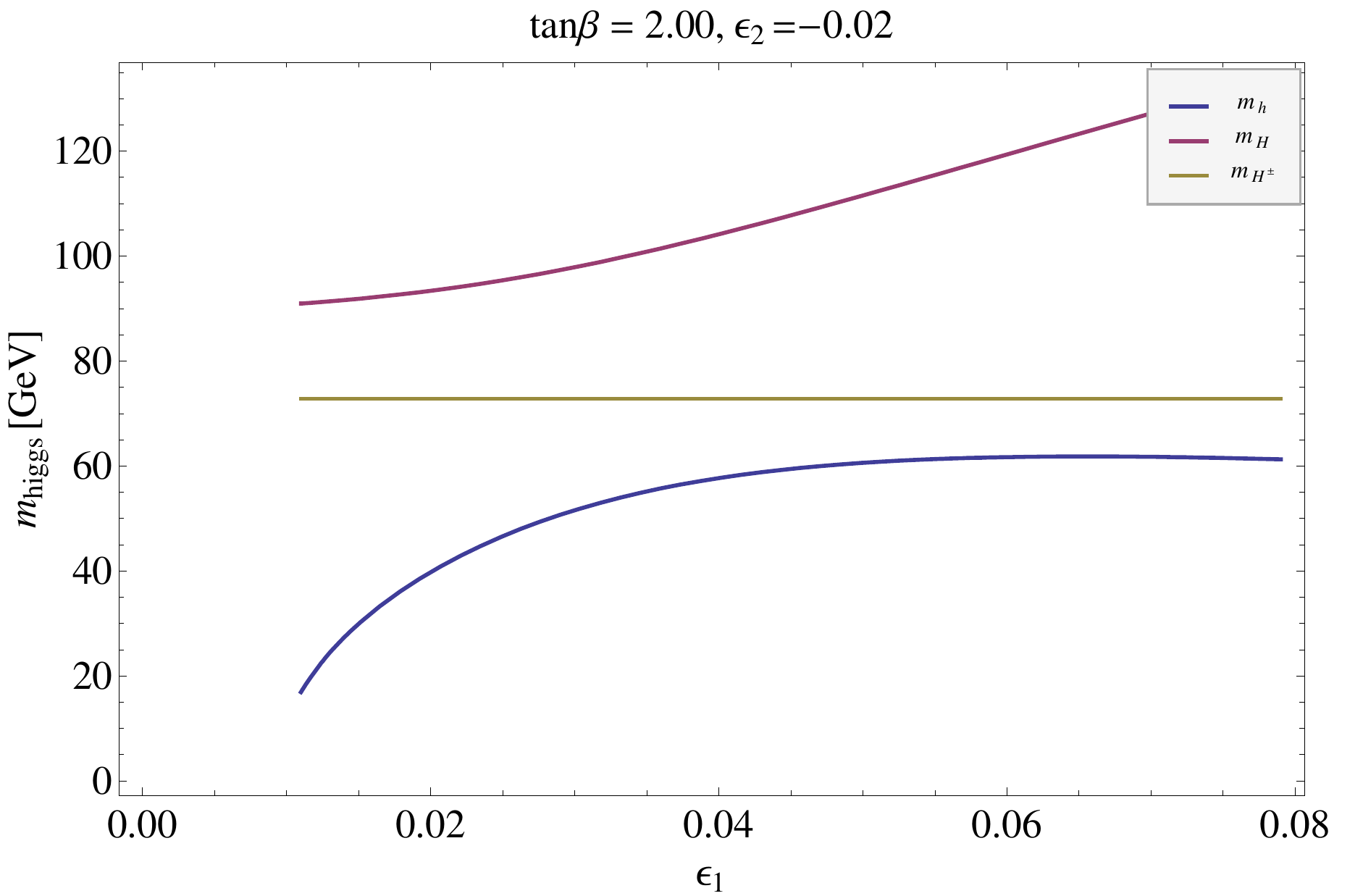}}
\subfigure[]{
\includegraphics[width=3.in]{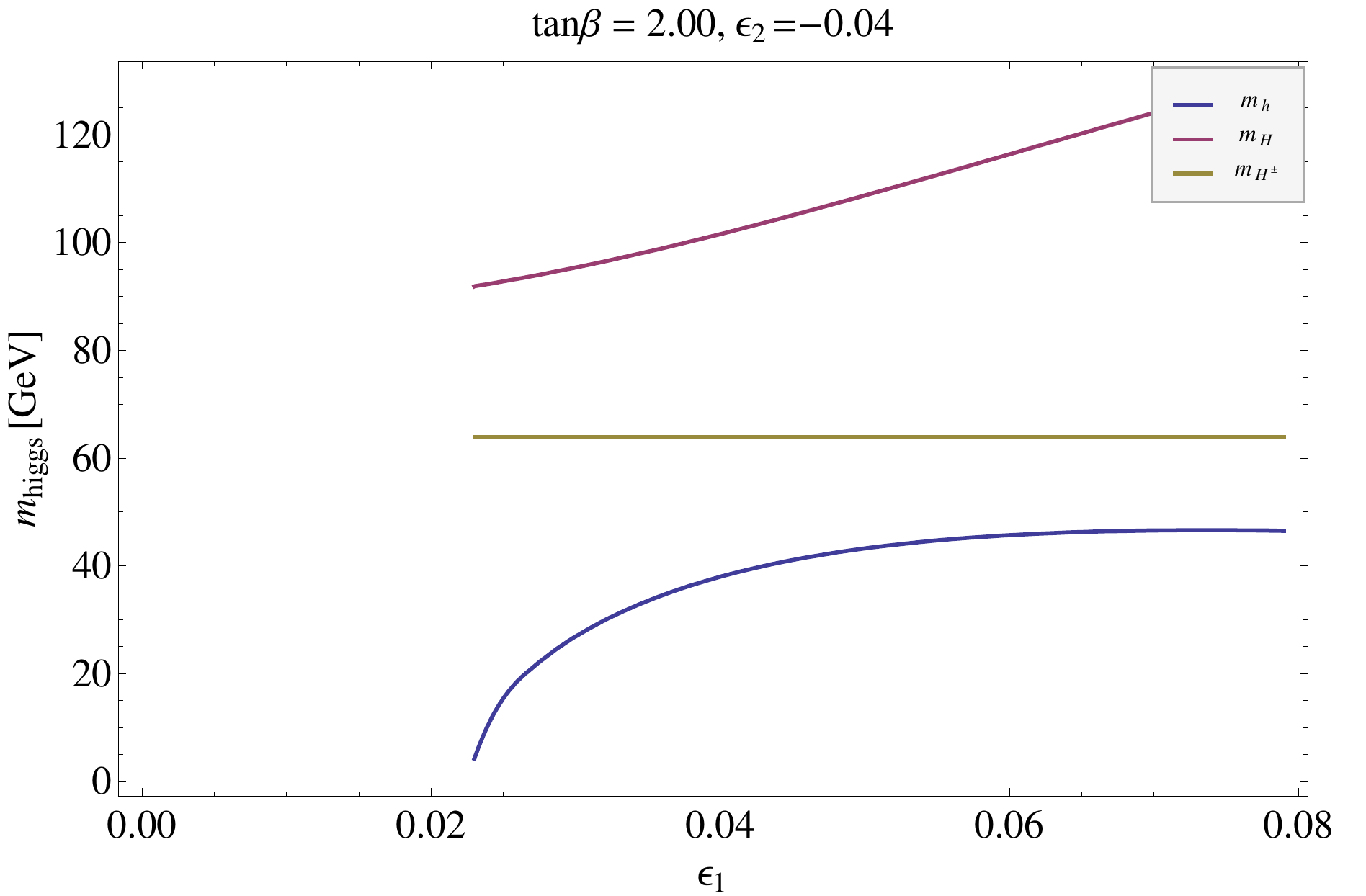}}
\caption{Higgs spectrum scanned for $\epsilon_1$ and $\epsilon_2$ for several $\tan\beta$. $\epsilon_3=0.01$, $\epsilon_4=0$ and $m_A=7$GeV are used here. Red, brown and blue lines denote $H, H^{\pm}$ and $h$, respectively.}
\label{fig:Hspectrum}
\end{figure}

Higgs spectrum is shown in Fig. \ref{fig:Hspectrum}. In the plots, the CP odd Higgs mass is chosen to be 7 GeV. A few GeV difference in $m_A$ makes a little difference in CP even and charged Higgs mass spectrum. Two representative values of $\tan \beta$ (1.04 and 2) are chosen. When $\tan \beta$ is chosen to 1.04 (close to 1) (Fig. \ref{fig:Hspectrum} (a)(b)(c)), the mass matrix of CP even Higgs has a special structure such that one of the eigenvalue is nearly constant ($M_Z$). The transition point appears when the off-diagonal element is almost canceled by $\epsilon_1$. In the following discussion, we are interested in the region in which the off-diagonal element is positive after all. ($\epsilon_1 > 0.04$ in the plots.)
 Thus to reduce the light CP even mass to 80 GeV, we choose $\epsilon_2$ to be negative.
 $\epsilon_2=-0.04$ provides the light CP even Higgs mass at around 80 GeV. However, in this case the charged Higgs turns out to be too light, 60 GeV, and the scenario is ruled out. When $\tan \beta=2$ (away from 1), the spectrum shows a very smooth transition
of the CP even Higgs mixing angle ($\alpha$) from negative values ($\alpha <0$)
to positive ones ($\alpha >0$).

%% CHECK LATER

The $hAA$ and $HAA$ couplings are modified with the presence of $\epsilon_1$, $\epsilon_2$ and $\epsilon_3$. ($\epsilon_4=0$ from now on.)
\bea
g_{hAA} & = & \lambda_0 \cos 2 \beta \sin (\beta+\alpha) \nn\\
&& -i v \left[ -2\epsilon_1 \cos 2\beta \sin(\beta-\alpha) - \epsilon_2 (3\cos(\beta-\alpha) - \cos 2\beta \cos(\beta+\alpha))+ 4 \epsilon_3 \cos^2\beta \sin\beta\cos\alpha \right], \nn \\
g_{HAA} & = & -\lambda_0 \cos 2 \beta \cos (\beta+\alpha) \nn \\
&&-i v \left[ 2\epsilon_1 \cos 2\beta \cos(\beta-\alpha)  - \epsilon_2 (3\sin(\beta-\alpha) + \cos 2\beta \sin(\beta+\alpha))+ 4 \epsilon_3 \cos^2\beta \sin\beta\sin\alpha \right] , \nn
\eea
where $\lambda_0 = -i M_Z^2/v$.

The corrections to the masses and couplings from $\epsilon_1$ and $\epsilon_2$ are also discussed in \cite{Berg:2009mq}. It is clear that $h \to AA$ dominates over $h \to b\bar{b}$ in most of parameter space. Nonetheless, there is a chance that $h \to AA$ can be suppressed compared to $h \to b \bar{b}$ since $h \to b \bar{b}$ is given by bottom Yukawa coupling alone (dominantly) while $h \to AA$ is given by two or three independent contributions. It is this cancelation which we will use to explain the forward backward asymmetry of bottom quark later in this paper. 

\begin{figure}[thb]
\subfigure[tan $\beta=1.04$]{
\includegraphics[width=3.in]{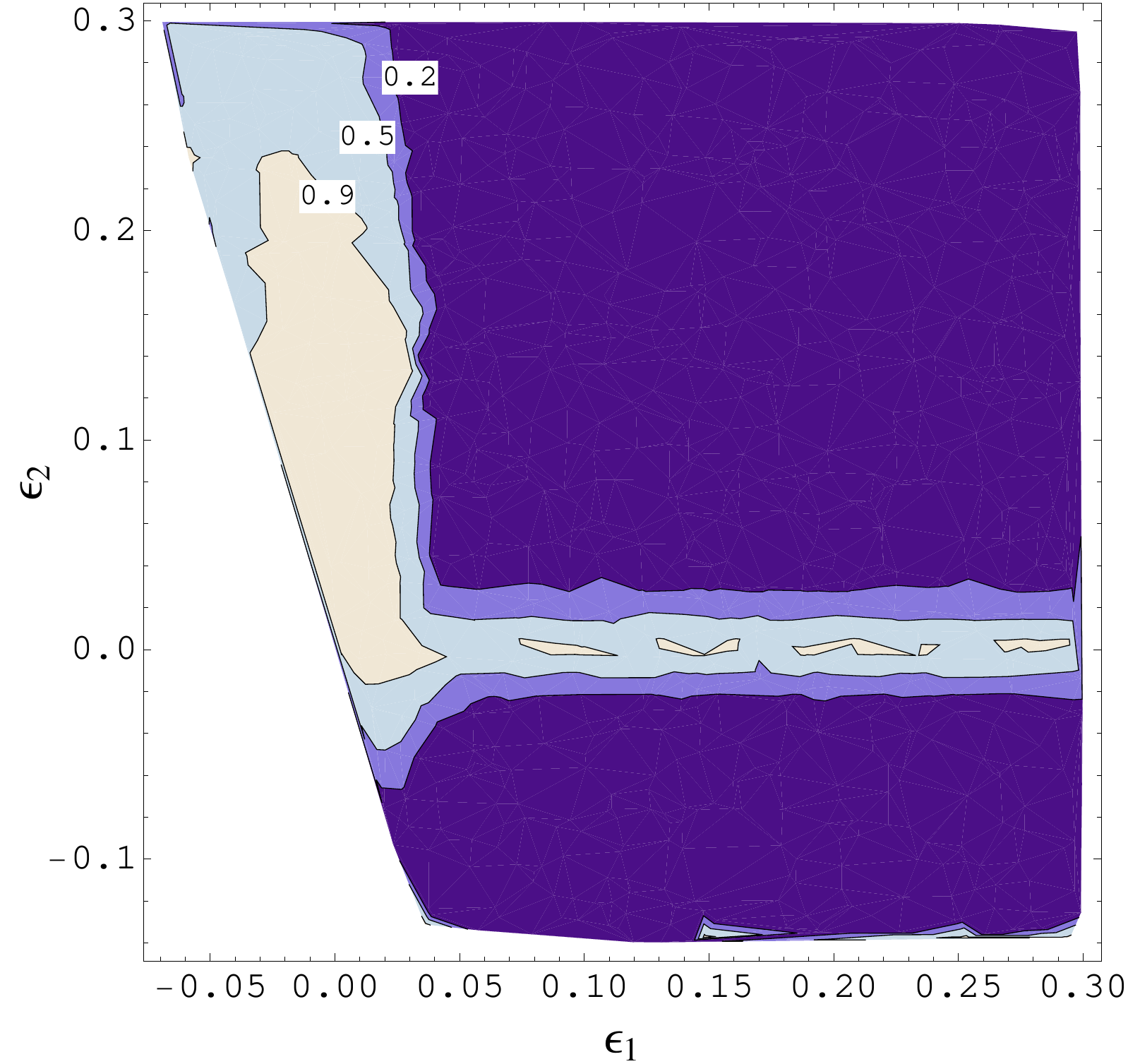}}
\subfigure[tan $\beta=1.6$]{
\includegraphics[width=3.in]{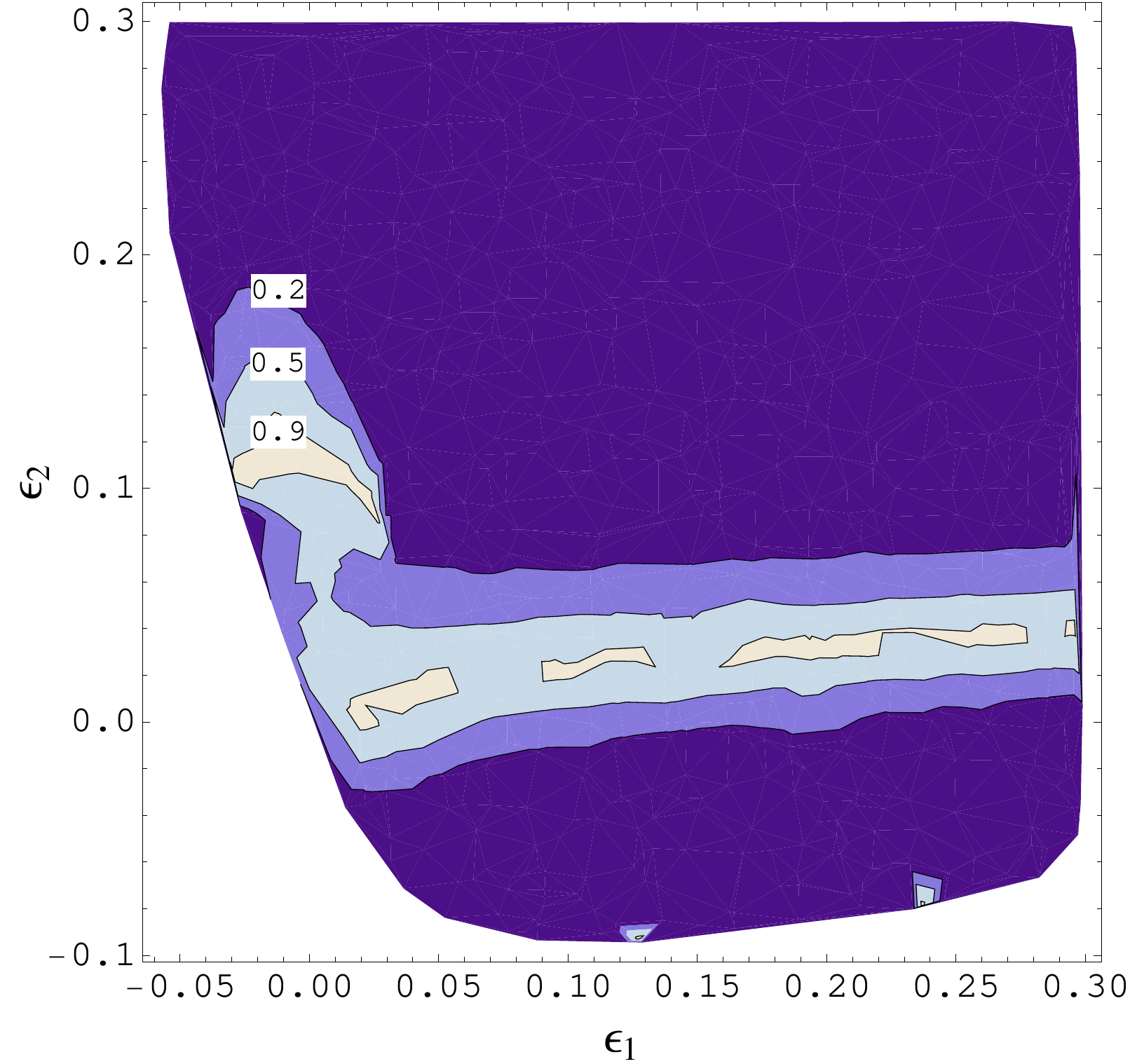}}
\subfigure[tan $\beta=2$]{
\includegraphics[width=3.in]{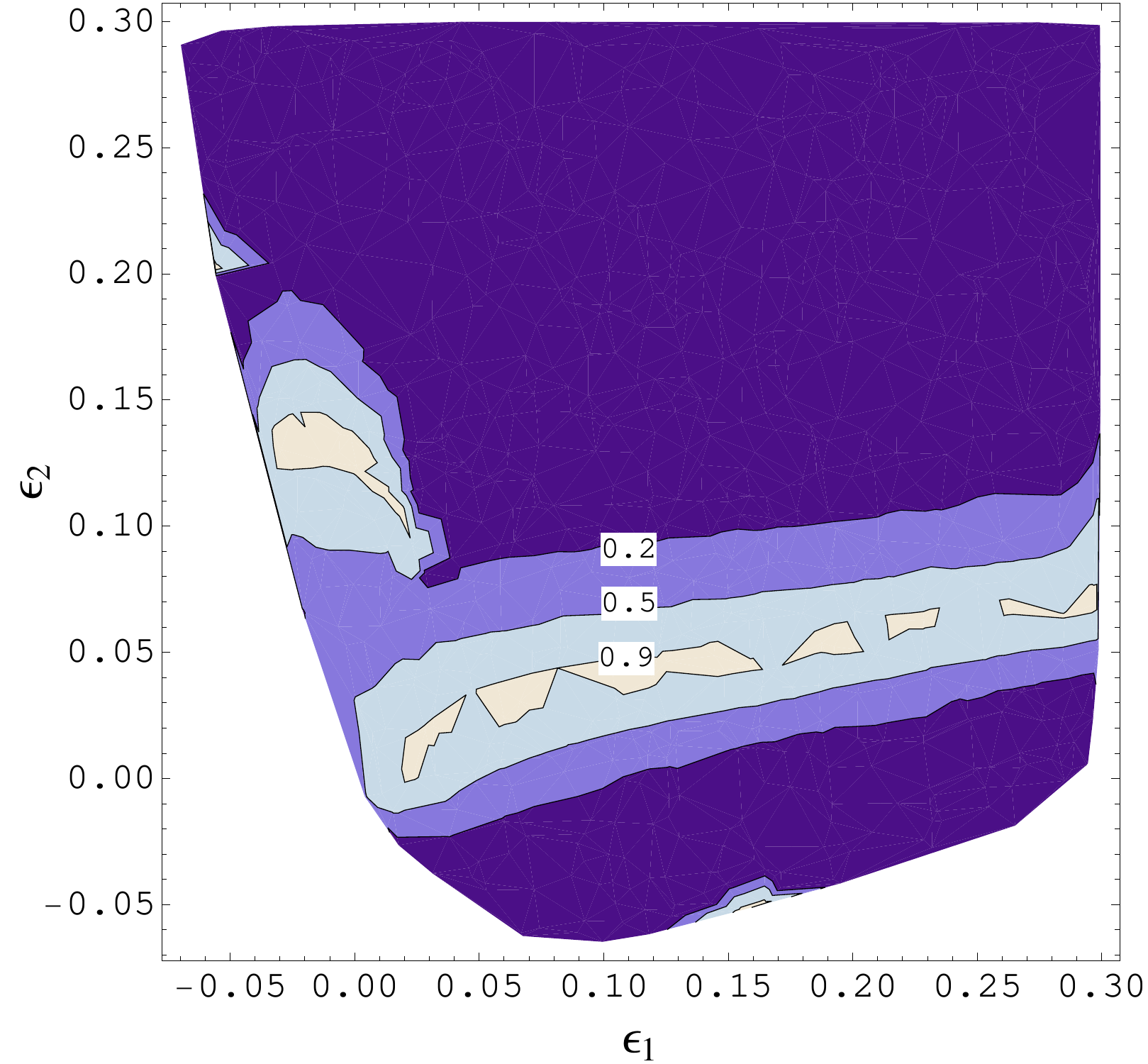}}
\caption{Branching ratio to $b\bar{b}$, $\frac{\Gamma(h \to b\bar{b})}{\Gamma(h \to b\bar{b})+\Gamma(h\to AA)}$ is plotted on $(\epsilon_1, \epsilon_2)$ space. $\epsilon_3=0.01$, $\epsilon_4=0$ and $m_A=7$GeV are taken here. At each point on the curves, Higgs mass may vary with respect to $\epsilon_{1,2}$.}
\label{fig:br}
\end{figure}

Fig. \ref{fig:br} shows the ratio of partial decay width $\Gamma(h \to b\bar{b})/(\Gamma(h \to b\bar{b})+\Gamma(h \to AA))$ for $\tan \beta = 1.04, 1.6$ and $2$, respectively. This is basically the branching ratio of $h \to b \bar{b}$ as $b\bar{b}$ and $AA$ provide dominant decay channels. ${\rm Br}(h \to b\bar{b})$ is larger than 50\% in the skyblue and ivory regions. 
When $\tan \beta=1.04$, it is possible to have a cancelation of $\epsilon_1$ and $\epsilon_2$ contribution to $g_{hAA}$ coupling and there appears a bulk region in between $\epsilon_1=0$
and $0.05$.  The horizontal line with $\epsilon_2 \sim 0$ in Fig. \ref{fig:br} (a) shows that $\epsilon_1$ dependence on $g_{hAA}$ almost vanishes since $\cos 2\beta \sim 0$.
In other plots, the cancelation appears as a line in $\epsilon_1$ and $\epsilon_2$ plane
and sizable ${\rm Br}(h \to b\bar{b})$ is possible along the line.

Assuming that the conditions is satisfied ($h \to b\bar{b}$ dominates over $h \to AA$),
we can compare how much of $b\bar{b}$ pairs can be produced from LEP compared to the ones directly produced from Z boson decay. ${\rm Br} (Z \to hA \to b\bar{b})$ depends on light CP even Higgs mass and CP odd Higgs mass because the most significant factor is the kinematic suppression at the Z pole. 
$A$ decays into soft 2$\tau$ or $c \bar{c}$ and we assume these soft jets are not tagged.

\begin{figure}[!ht]
\resizebox{0.6\columnwidth}{!}
{\includegraphics[height=60mm]{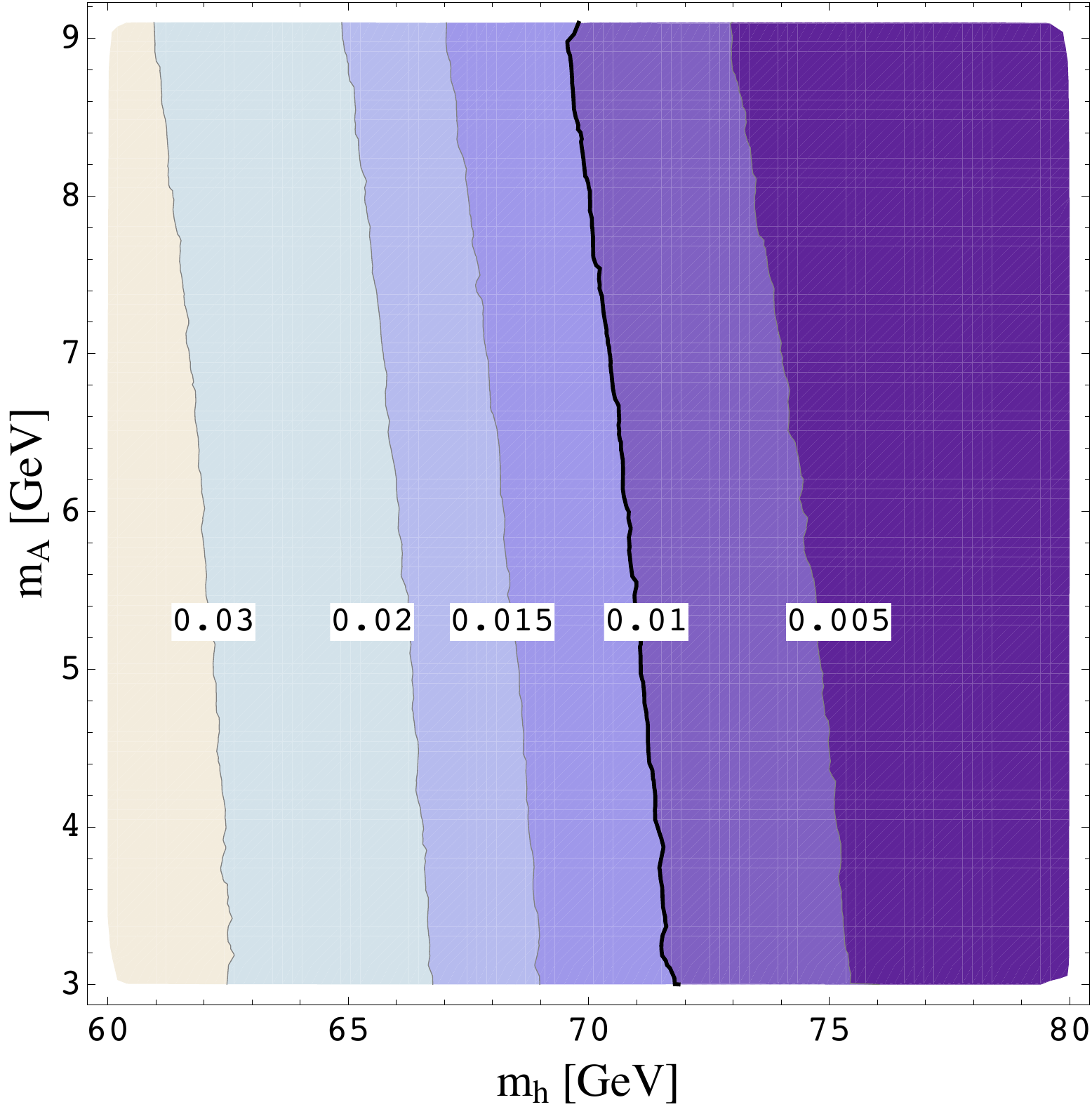}}
\caption{The ratio of bottom quark pair produced from Higgs vs Z, $\sigma (e^+e^-\rightarrow hA\rightarrow b\bar{b})/\sigma (e^+e^-\rightarrow Z\rightarrow b\bar{b})$, is plotted on $(m_h,m_A)$ space.}
\label{fig:BR}
\end{figure}

Fig. \ref{fig:BR} shows the ratio of bottom quark events produced from Higgs compared to those from Z. Here $Z \to Zh$ is assumed to be zero and $Z \to hA$ coupling is maximized. Also Br($h \to AA$) is taken to be 1.
For $m_h=75$ GeV, the ratio can be about 0.5 \% and if it is lighter, $m_h=70$ GeV, it can be as large as 1\%. The direct search limit from the LEP allows the light CP even Higgs as light as 72 GeV\cite{LHWG:2005}.

In this section it is shown shown that the new parameter space of BMSSM allows light Higgs scenario. CP odd Higgs mass is about 7 or 8 GeV (lighter than $2m_b$) and light CP even Higgs is about 70 to 75 GeV which avoiding model independent search bound from OPAL by having small $ZZh$ coupling. There are two consequences of this scenario.
One is that on-shell $hA$ production is possible at the Z pole and there is a possibility that we are faked to believe that all the bottom quark pairs are from Z decay even in the case $h$ decays into $b\bar{b}$.
This might cause an immediate contradiction with the observation of $R_b$ consistent with the SM prediction. 
However, it remains as a perfect possibility here since the other consequence of the scenario can reconcile it.
The scenario predicts a very light charged Higgs, $m_{H^\pm} \sim m_W$.
For such a light charged Higgs, $Z \to b\bar{b}$ vertex gets a correction of order 1\%
and it is consistent with new contributions from $Z \to hA \to b\bar{b}$.
Also such a light charged Higgs is not ruled out by current experimental bounds
since $H^{\pm} \to W^* A$ can have a sizable branching ratio while the standard bound is given
with the assumption that $H^{\pm} \to \tau \nu$ (or $H^{\pm} \to c \bar{s}$) is the dominant decay mode
\cite{Dermisek:2008dq} \cite{Park:2006gk}.
In the following sectioin we review the status of LEP electroweak precision data
and discuss the implication of this new scenario on LEP data.
All the discussion in this section can also be considered in the context of two Higgs doublet model (2HDM) as 2HDM is a generalization of BMSSM as long as Higgs sector is concerned.

\section{LEP Electroweak Precision Data : $A^b_{FB}$ and $R_b$}

LEP experiment is known as one of the most successful ones which confirm the Standard Model (SM) at a very high precision, one per mil. This acts as the main source of the frustration for any physics beyond the SM. In this paper we take a viewpoint that 3 sigma deviation of $A^b_{\rm FB}$ between the prediction and the measurement existing in LEP data might be a hint for a new physics.
We discuss whether light Higgs scenario might be relevant to reduce the tension.

Forward-backward asymmetry of bottom quark produced from electron-positron pair,
\bea
A^b_{\rm FB} (M_Z) 
& = & \frac{\sigma_F - \sigma_B}{\sigma_F + \sigma_B}  =  \frac{3}{4} A_e A_b \\
& = & \frac{3}{4} \frac{((g^e_L)^2-(g^e_R)^2)}{((g^e_L)^2+(g^e_R)^2)} \frac{((g^b_L)^2-(g^b_R)^2)}{((g^b_L)^2+(g^b_R)^2)}. 
\eea
has been measured at LEP I and LEP II
and it is the electroweak observable which shows the largest discrepancy compared to the standard model prediction:
\bea
&A^b_{\rm FB} & =  0.0992 \pm 0.0016,  \\
&A^b_{\rm FB\ SM} & =  0.1037 \pm 0.0008. 
\eea

The difference $0.0045$ corresponds to $2.8 \sigma$ deviation with the experimental error
or $2.5 \sigma$ deviation with the combined error.
The strong constraints on the bottom quark pair production at the Z pole from $R_b$ measurements
makes hard to resolve this deviation.

\bea
R_b & = & \frac{\Gamma (Z \to b \bar{b})}{\Gamma(Z \to {\rm hadrons})} 
= \frac{((g^b_L)^2+(g^b_R)^2)}{\sum_{q} ((g^q_L)^2+(g^q_R)^2)}, 
\eea
where $q$ represents five quarks excluding top quark.
The observed value and the standard model prediction agrees well:
\bea
&R_b & = 0.21629 \pm 0.00066,  \\
&R_b^{\rm \ SM} & = 0.2158. 
\eea
Now the difference $0.0005$ corresponds to $0.7 \sigma$.

The discrepancy of $R_b$ is $0.3 \%$,
but the difference in $A^b_{\rm FB}$ is about $4.4\%$
- which is more than ten times larger than that of $R_b$.
Any new physics responsible for the deviation of $A^b_{\rm FB}$ should preserve 
the branching ratio of $Z \to b \bar{b}$ with a high precision.
If $g^b_L$ and $g^b_R$ were comparable in size, 
the total sum could be preserved by modifying $g^b_L$ and $g^b_R$ at a percent level with the opposite direction.
However, in reality, the tree level value is
\bea
g^b_L & = & -\frac{1}{2}+\frac{1}{3} \sin^2 \theta_W \simeq -0.42, \\ 
g^b_R & = & \frac{1}{3} \sin^2 \theta_W \simeq 0.08, 
\eea
where the left-handed contribution is about 27 times larger.
In general it is possible for a left-handed coupling 
to obtain a percent level correction from a loop but to keep the branching ratio to be the same,
we need about 25 percent correction to right-handed coupling $g^b_R$.
Without significant modification of the right-handed coupling, 
the loop correction to the left-handed coupling is typically negative (from charged Higgs) 
and make it impossible to keep $R_b$ as it is.

If the charged Higgs is very light (lighter than 100 GeV) and $\tan \beta$ is close to 1, 
the (B)MSSM correction to $Z \to b\bar{b}$ can be sizable such that $R_b^{\rm SM}$ can be reduced
by 0.5\%, 1\% or 1.5\% \cite{Boulware:1991vp}.

\begin{figure}[!ht]
\resizebox{0.6\columnwidth}{!}
{\includegraphics[height=60mm]{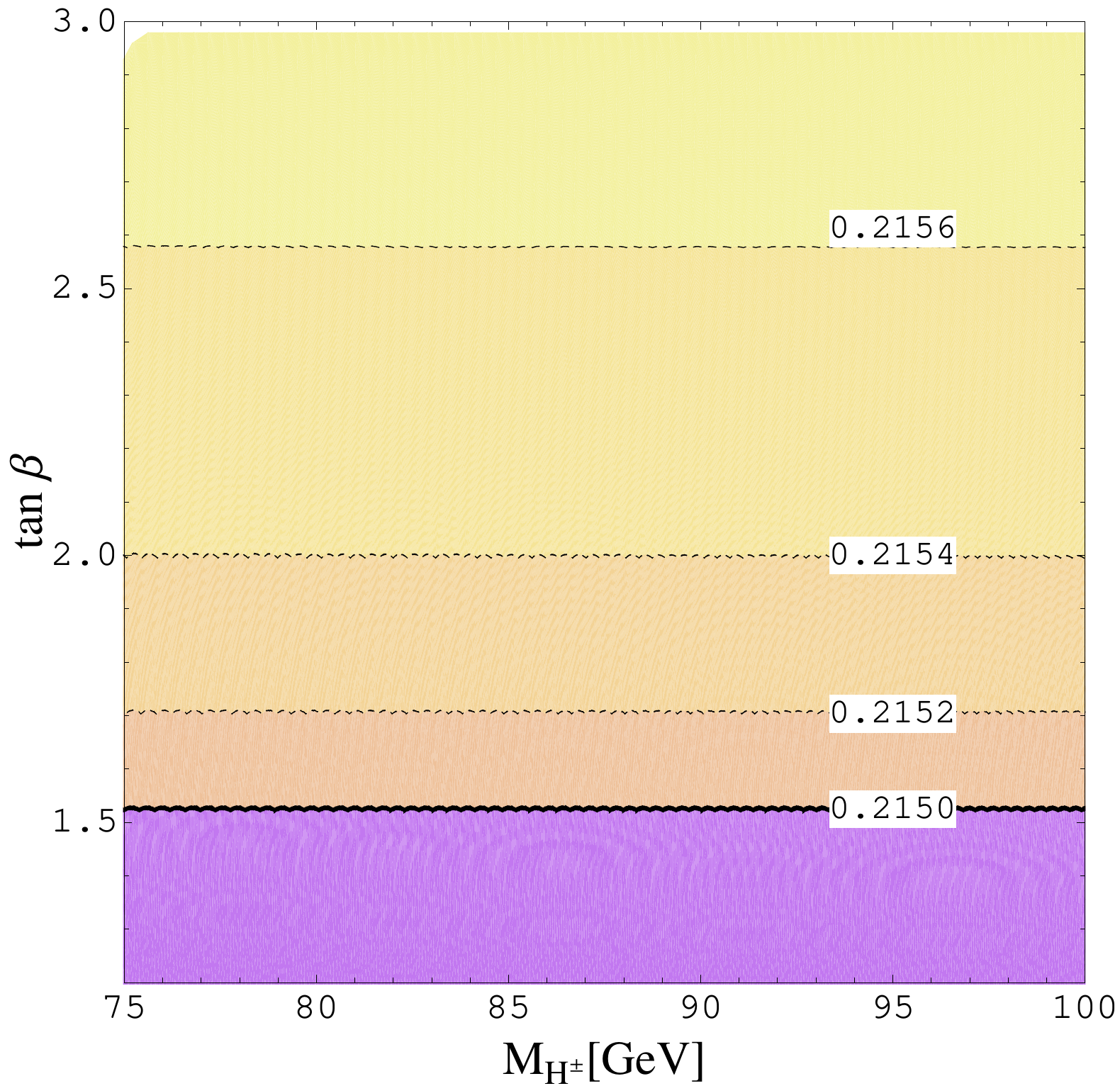}}
\caption{$R_b^{\rm BMSSM}$ values from loop correction of charged Higgs and SUSY. $m_{\tilde{q}_3}=m_{\tilde{t}}=190$GeV, $A_t =\mu =200$GeV. This is calculated by our own code using formulae of \cite{Boulware:1991vp}.}
\label{fig:Rb}
\end{figure}

Light supersymmetric particles can cancel the charged Higgs loop.
Fig. \ref{fig:Rb} shows $R_b^{\rm BMSSM}$ which also includes light stop.
For $\tan \beta=1.5$, the predicted $R_b^{\rm BMSSM}=0.2150$
which is about $0.7$ \% smaller than the measured value.

The correction appears in $g_L$ and it is impossible to make a sizable correction to $g_R$ (27 times larger one compared to $g_L$)
from loop corrections
to keep $R_b$ as it is.
Such modification of $g_L$, $g_R$ is not required 
if there are new processes which gives fake $Z \to b \bar{b}$ signals.
In light Higgs scenario of BMSSM,
$Z \to hA \to 2b + (2\tau \;$ or $\; c\bar{c})$ can be counted as $Z \to 2b \;+ \; gluons$ events 
when tau jets are soft,
because it is hard to distinguish soft tau jets from QCD backgrounds.

In this case, measured $R_b$ value should be compared to $R_b^{\rm BMSSM\ total} = R_b^{\rm BMSSM} + R_b^{\rm fake}$
and measured $A^b_{\rm FB}$ becomes
\bea
A^b_{\rm FB}
& = & \frac{\sigma_F - \sigma_B}{\sigma_F + \sigma_B + \sigma_{fake}}.
\eea
Because Higgs is a scalar particle, it evenly contributes to $\sigma_F$ and $\sigma_B$, 
so it decreases $A^b_{\rm FB}$.
BMSSM has a parameter space such that prediction of $R_b$ is sufficiently smaller than $R_b^{SM}$
so $\sigma_{\rm fake}$ ($R_b^{\rm fake}$) can be large enough,
a sizable loop correction to $g^b_R$ is not required.

However, it is not clear how many of the events will be included in data set for $A^b_{\rm FB}$.
To measure $A^b_{\rm FB}$, the direction of bottom quarks is required.
Events which contain hard initial state radiation(ISR) photons or final state radiation(FSR) gluons 
should be removed because it disturbs the direction of beam or bottom quarks.
For example, OPAL uses combination of
sphericity(\cite{Acton:1992sc}), 
total energy, 
energy imbalance along the beam direction(\cite{Alexander:1991qw})
and so on, to select such hadronic decay events.
In $Z \to hA \to 2b (2\tau \;$ or $\; c\bar{c})$ events, 
angle between $\tau$($c$) and $b$ can be large because two jets are loosely correlated.
In this case, tau jets can look like hard gluon jets from b-quarks,
so it might be rejected by the selection cut of $A^b_{\rm FB}$.

To make the discussion simple, we assume ${\rm Br}(h \to b\bar{b})=1$ and ${\rm Br}(A \to \tau^+ \tau^-)=1$.
Also we assume that all the events of $Z \to hA$ are recorded as $Z \to b\bar{b}$
due to the softness of tau jets.  
The exact fraction of events which is included in $A^b_{\rm FB}$ needs a detailed information
of the events. Here we consider the most optimistic case in which 
all the events generated from $Z \to hA$ are included in $A^b_{\rm FB}$ measurement.
Thus the estimation here would work as the maximum correction 
we can expect from this scenario.

\bea
R_b & = & \frac{\Gamma (Z \to b \bar{b})+\Gamma(Z\to hA)}{\Gamma(Z \to {\rm hadrons})+\Gamma(Z\to hA)}. 
\eea

Let $a=\frac{\Gamma(Z\to hA)}{\Gamma (Z \to b \bar{b})}$. For very small $a \ll 1$,
\bea
R_b & = & \frac{\Gamma (Z \to b \bar{b})}{\Gamma(Z \to {\rm hadrons})}
\left(1+(1-R_b)a \right). 
\eea
Therefore, if one loop correction decreases $(g_L^b)^2$ by one percent,
$R_b$ remains to be the same if $a=0.012$. 
When there is no change in $(g_L^b)^2$, $a=0.012$ will increase $R_b$ by one percent
which is the correction allowed by 2$\sigma$ of $R_b$ measurement.
Total Z width is precisely measured and should be kept within $0.1\%$.
$a=0.012$ gives $0.15\%$ increasement of total Z width which is in $1.5\sigma$ range.

The same $a$ enters in $A^b_{\rm FB}$ ($A^b_{\rm FB}=\frac{3}{4} A_e A_b$).
\bea
A_b & = & \frac{\Gamma(Z \to b_L \bar{b_L})-\Gamma(Z \to b_R \bar{b_R})}{\Gamma(Z \to b_L \bar{b_L})+\Gamma(Z \to b_R \bar{b_R})+ \Gamma(Z \to hA)}  \\
& = & \left(\frac{\Gamma(Z \to b_L \bar{b_L})-\Gamma(Z \to b_R \bar{b_R})}{\Gamma(Z \to b_L \bar{b_L})+\Gamma(Z \to b_R \bar{b_R})}\right) \left( 1-a \right). 
\eea
Therefore, $a=0.012$ will reduce $A_b$ (and thus $A^b_{\rm FB}$) by $1.2\%$
which will make the difference between the prediction to be 0.0033 which is within 2 $\sigma$ (1.8 $\sigma$).  Here we simply assumed that $h A$ production is isotropic and appears only in the denominator and cancels in the numerator. Whether $1-a$ is the right dependence depends on the fitting method of angular distribution. The actual change to $A_b$ from $a$ can be smaller than what is estimated here.
Note that changing $g^b_L$ can not change $A_b$ very much since it dominates in the numerator and denominator at the same time.

\section{Flavor Changing Neutral Currents : $b \to s \gamma$}

It might be challenging to explain $b \to s \gamma$ with $m_H^{\pm} = 80$ GeV
as the lower mass bound on the charged Higgs is about $250$ GeV
if only the charged Higgs contribution is taken into account.
In the MSSM, light stop is also natural and can be responsible for the cancelation of charged Higgs contribution.
2 $\sigma$ range of $b \to s \gamma$ is obtained from the cancelation of the charged Higgs loop with the stop-chargino loop.

\begin{figure}[!ht]
\resizebox{0.6\columnwidth}{!}
{\includegraphics[height=60mm]{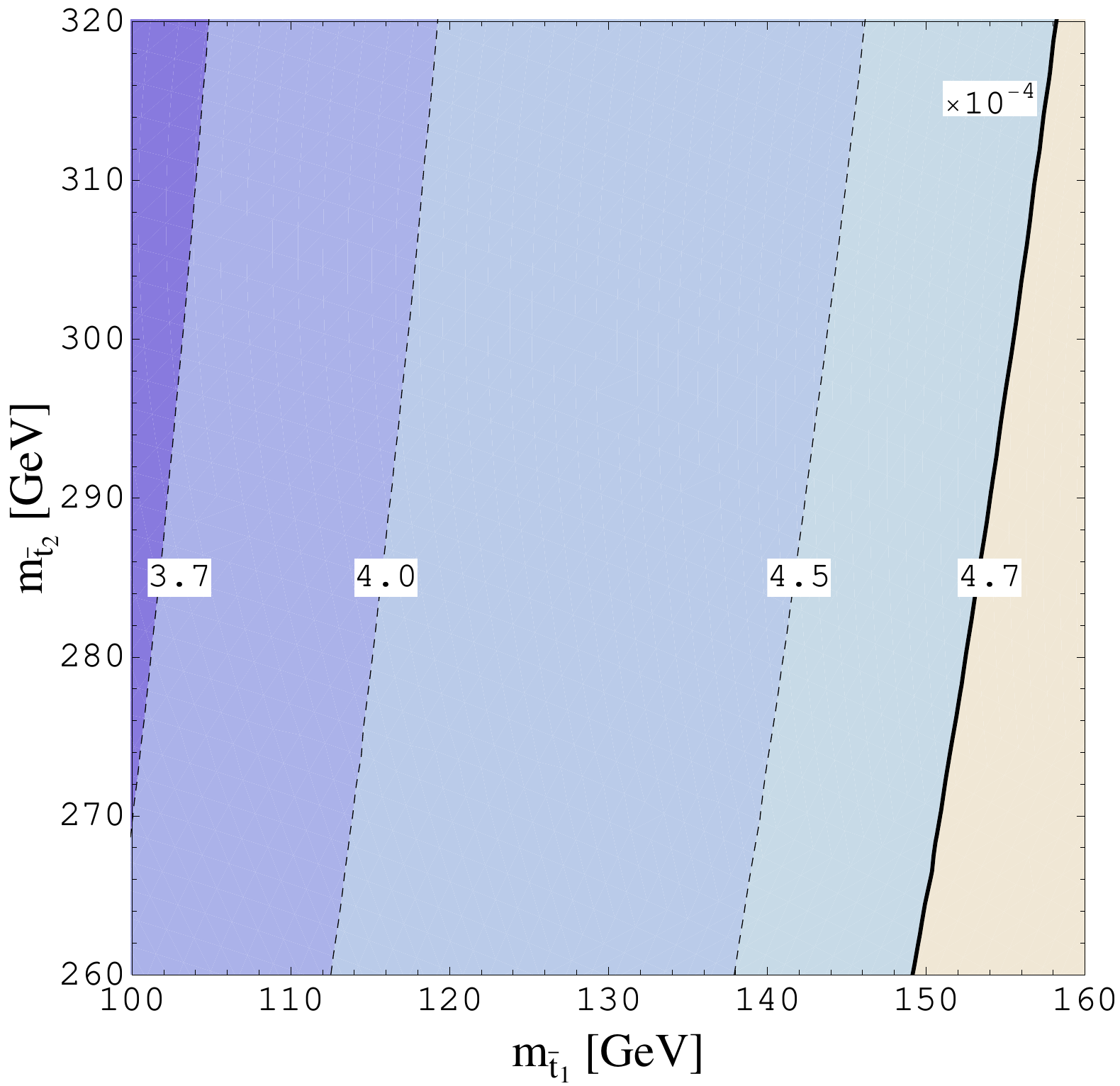}}
\caption{This is contour plot of Br($b\to s\gamma$) for $M_2 =200$GeV, $\epsilon_1 =0.0096$, $\epsilon_2=0.00086$, $2\epsilon_3 v^2 \sin^2\beta =708 \text{GeV}^2$ and $\tan\beta =1.9$. This is calculated by SuperIso v2.3\cite{Mahmoudi:2008tp}}
\label{fig:bsg}
\end{figure}

At first glance, $b\to s \gamma$ observable is not dramatically changed in BMSSM framework. The reason is that Higgs sector, especially Higgs cubic and quartic couplings, suffers from considerable modification but chargino and squark sectors do not. However, such BMSSM operators can modify higgsino operators and then modify chargino and neutralino mixing. Futhermore, squark mixing is also changed by superpotential modification by $\epsilon_1$ operator. In most natural case, stop and chargino masses are also as light as Higgs masses, and small modification to Higgsino sector could be important for $b\to s \gamma$ observable. BMSSM modification to chargino sector is given in chargino mass matrix
\begin{equation}
X=
\begin{pmatrix}
M_2 & \sqrt{2}M_W \sin \beta\\ \sqrt{2}M_W \cos \beta & \mu
\end{pmatrix}
+ \frac{\epsilon_1}{\mu}v^2 \sin 2\beta
\begin{pmatrix}
0 & 0 \\ 0 & 1
\end{pmatrix}.
\end{equation}
Stop mass matrix is given by
\begin{equation}
m_{\tilde{t}}^2 =
\begin{pmatrix}
m_{Q_3}^2+m_t^2+\Delta_u & m_t(A_t - \mu \cot \beta) \\ m_t(A_t - \mu \cot \beta) & m_{\bar{u}_3}^2+m_t^2+\Delta_{\bar{u}}
\end{pmatrix}
-2m_t\frac{\epsilon_1}{\mu}v^2\cos^2 \beta
\begin{pmatrix}
0 & 1 \\ 1 & 0
\end{pmatrix}.
\end{equation}
Consequently, chargino and stop mixing matrices are also changed.
$\epsilon_1$ term changes supersymmetric $\mu$ term as
\begin{equation}
\mu \to \mu+\frac{\epsilon_1}{\mu}v^2\sin 2\beta.
\end{equation}
Therefore, we can have effectively large higgsino mass and large stop mixing for small $\tan \beta$ case in which we are interested.
If $M_2 \sim \mu \sim 100$GeV and $\epsilon_1 \sim 0.1$, 
supersymmetric $\mu$ term can be enhanced 20-30\% for $1\lesssim \tan\beta \lesssim 2$ 
and then $\epsilon_1$ contribution could modify $b\to s \gamma$ observable. Numerical result is shown in fig. \ref{fig:bsg}. 

The thick black line at $4.7 \times 10^{-4}$ is the 2 $\sigma$ bound. Only the left part from the thick black line is consistent with the measured branching ratio of $b \to s\gamma$. In this case the lightest stop mass should be less than 160 GeV which is certainly lighter than the top quark. It is generally the case for different choice of $\tan \beta$ and other supersymmetry breaking parameters. Wino/higgsino mass also should be light since large loop corrections due to small $\tan \beta$ and light charged Higgs have to be canceled by supersymmetric counter terms (stop-chargino loop) which can be sizable when stop and chargino are light enough. This provides an interesting connection between our scenario and the electroweak baryogenesis which works only when the right-handed stop is lighter than top quark.
The detailed study of the implication of BMSSM operators to $b \to s \gamma$ and muon $g-2$ will be given elsewhere \cite{Bae:2009}.

\section{Conclusion}

In this paper we explored the phenomenology of light Higgs scenario in BMSSM.
More specifically, we found that h and A production at the Z pole is kinematically allowed
in BMSSM.
For very light A ($\le 10$ GeV)  and h slightly lighter than Z ($\sim 70$ GeV)
which is still consistent with the LEP data,
h and A are produced with small momentum.
Bottom quark pairs produced from h are close to back to back
and tau pairs from A are soft enough.
Forward backward asymmetry or $R_b$ measurement at LEP
can be affected by bottom quarks produced from Higgs.
For $\tan \beta \sim 1.5$ to $2$, it is possible to find a parameter space
which can ameliorate the tension in $A^b_{\rm FB}$
without spoiling nice agreement in $R_b$.

This scenario at the same time predicts very light charged Higgs $m_{H^\pm} \sim m_W$
but still it can be consistent with current $b \to s \gamma$ observation
by having a cancelation with light stop-chargino loop.
To confirm whether this is the case or not, more careful analysis
on discriminating soft taus from gluon jets and angular distribution fit is needed.
Independently of its relation to $A^b_{\rm FB}$,
light CP odd Higgs scenario in BMSSM or 2HDM is interesting enough
and further exploration is left for future work.

\section*{Acknowledgement}

HDK thanks Nima Arkani-Hamed for interesting discussions on the topic.
This work is supported by KRF-2008-313-C00162 (KJB, DK, HDK, JHK)
and NRF with CQUEST grant 2005-0049049 (DK, HDK).

\end{document}